\documentclass{basi}
\usepackage[english]{babel}
\usepackage[varg]{txfonts}
\begin{document}
\title{TIRCAM2: The TIFR Near Infrared Imaging Camera}
\author[M.B. Naik et al]{M.B. Naik, D.K. Ojha$\thanks{email: ojha@tifr.res.in}$, S.K. Ghosh, S.S. Poojary, R.B. Jadhav,
\newauthor G.S. Meshram, P.R. Sandimani, S.B. Bhagat, S.L.A. D'Costa,
\newauthor S.M. Gharat, C.B. Bakalkar, J.P. Ninan and J.S. Joshi
\\ \\
Infrared Astronomy Group, Department of Astronomy \& Astrophysics, \\ Tata 
Institute of Fundamental Research, Homi Bhabha Road, Colaba, Mumbai - 400 005, India}

\pubyear{2012}
\volume{00}
\pagerange{\pageref{firstpage}--\pageref{lastpage}}

\date{Received --- ; Accepted --- }

\maketitle

\label{firstpage}

\begin{abstract}
TIRCAM2 (TIFR Near Infrared Imaging Camera - II) is a closed cycle cooled imager 
that has been developed by the Infrared Astronomy Group at the Tata Institute
of Fundamental Research  for observations in the near infrared band of
1 to 3.7 $\mu$m with existing Indian telescopes. In this paper, we describe
some of the technical details of TIRCAM2 and report its observing capabilities, 
measured performance and limiting magnitudes with the 2-m IUCAA Girawali telescope 
and the 1.2-m PRL Gurushikhar telescope. The main highlight is the camera's 
capability of observing in the nbL (3.59 $\mu$m) band enabling our primary 
motivation of mapping of Polycyclic Aromatic Hydrocarbon (PAH) emission 
at 3.3 $\mu$m. 
\end{abstract}


\section{Introduction}

TIFR had developed a near infrared (NIR) imaging camera, named  TIRCAM1 
(TIFR Near Infrared Imaging Camera - I), for astronomical imaging applications, 
which was based on a SBRC InSb focal plane array (FPA) (58 x 62 pixels),
sensitive between 1 and 5 $\mu$m. Astronomical observations with \mbox{TIRCAM1} 
were regularly carried out during 2001 - 2006 at the f/13 Cassegrain focus of 
the Mount Abu 1.2-m telescope belonging to Physical Research Laboratory (PRL), India. 
The TIRCAM1 system and related work have been described in Ghosh \& Naik (1993), 
Ghosh (2005) and Ojha et al. (2002, 2003, 2006). 

TIRCAM1 has now been upgraded to TIRCAM2 with the aim of realising a larger 
format detector array for use with the 2-m
Himalayan {\it Chandra} Telescope (HCT) at Hanle (Ladakh, India),   
to utilize its full capability in the range of 1 to 3.7 $\mu$m and, with a
few modifications in the optics, the entire range of 1 to 5 $\mu$m.  
TIRCAM2 uses a Raytheon InSb FPA (Aladdin III Quadrant 512 x 512 pixels) 
and f/9 re-imaging lens system optimized for observations with the HCT.  
The pixel size is 27 $\mu$m square. The quantum 
efficiency (QE) of the FPA is greater than 80\% from 1 to 5 $\mu$m. The TIRCAM2 system has seven 
filters and one block disk. The FPA operates at 35 deg K and is cooled by a closed 
cycle cooler. The engineering tests and science observations with TIRCAM2 
were carried out successfully at IUCAA Girawali Observatory (IGO), $\sim$ 80 km from Pune (India), with 
the engineering as well as the astronomy grade FPAs. 
A science run was also conducted at the PRL Mount Abu telescope. 

In this paper, we describe some of the technical details of the TIRCAM2 system
and a sample of the astronomical observations are presented to illustrate the performance characteristics.

\section{Subsystems of TIRCAM2}

\subsection{Dewar}  

\begin{figure}[ht]
\begin{minipage}[b]{0.5\linewidth}
\centering
\includegraphics [scale=1] {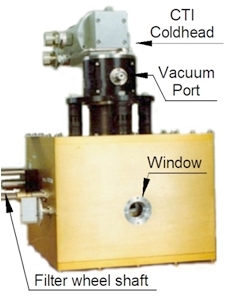}
\caption{A photograph of the \mbox{TIRCAM2} dewar system.}
\label{fig:dewarphoto}
\end{minipage}
\hspace{0.5cm}
\begin{minipage}[b]{0.5\linewidth}
\centering
\includegraphics [scale=0.4] {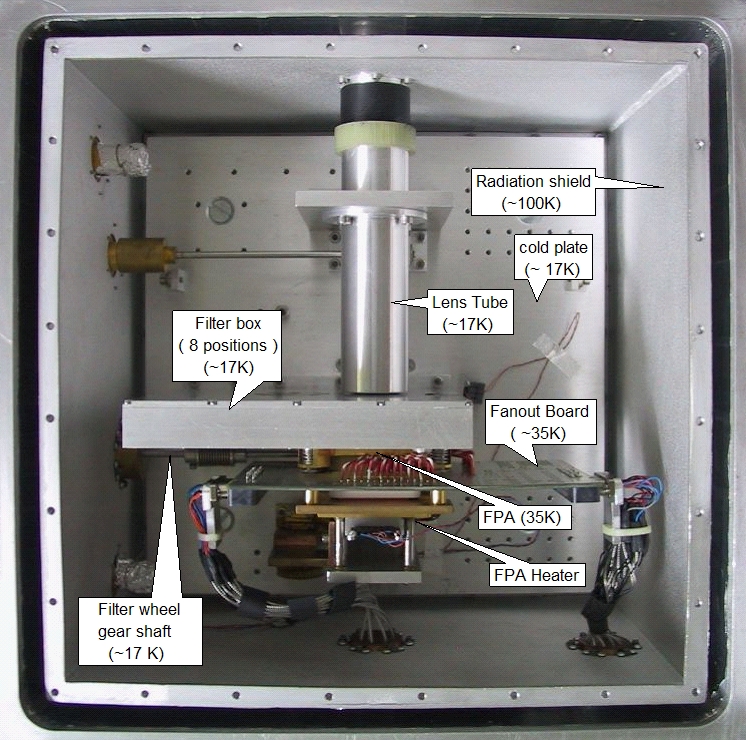}
\caption{Inside view of the \mbox{TIRCAM2} dewar system.}
\label{fig:dewarintphoto}
\end{minipage}
\end{figure}

\begin{figure}[ht]
\centering
\includegraphics [scale=0.6] {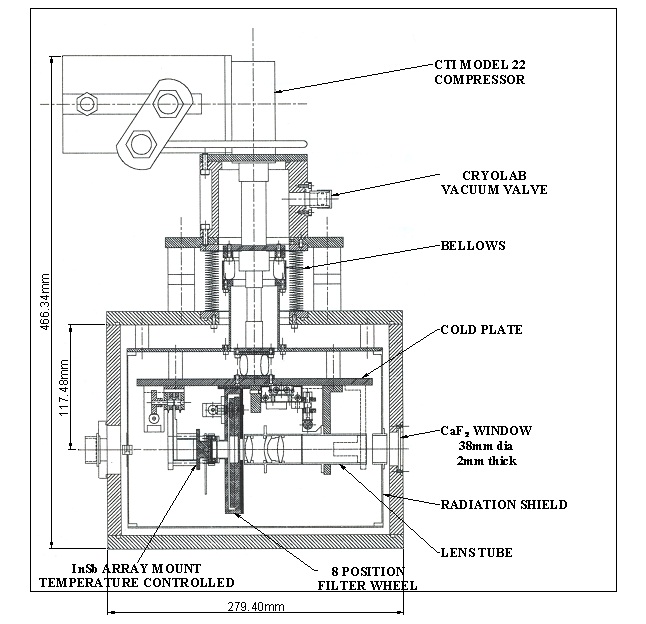}
\caption{A schematic of the TIRCAM2 dewar system.}
\label{fig:dewarschematic}
\end{figure}

A photograph of the TIRCAM2 dewar is shown in Fig. \ref{fig:dewarphoto} and the
inside view of the dewar and the schematic are shown in Figs. \ref{fig:dewarintphoto} and
\ref{fig:dewarschematic} respectively. The dewar has been manufactured by M/s. Infrared
Laboratories, USA. It consists of a vacuum jacket with a cold plate inside,
maintained at about 17 deg K by the second stage of a closed cycle cryocooler,
 on which are mounted the detector, filters and optical components
which are to be cooled (see Fig. \ref{fig:dewarintphoto}). The cold plate is surrounded by
a radiation shield which is maintained
at about 100 deg K by the first stage of the two stage closed cycle cryocooler.
The window is made of $CaF_{\rm2}$ through which the telescope beam enters
the photometer. On the side adjacent to the window, the filter wheel shaft is mounted using
a ferrofluidic feedthrough which allows the filter wheel to be rotated without affecting the
vacuum inside the dewar. The filter wheel shaft is connected to a stepper motor which rotates
the filter wheel to the desired position. At the top is a CTI make cold head which is a 
part of the CTI Model 22 cryocooler system. The cold head has two ports for gas inlet \& outlet 
and an electrical connector to drive its internal coldhead motor. Below the cold head is 
a vacuum port for evacuation of the dewar. The cold head vibrations are damped by the bellow 
and rubber bushes. The schematic of the dewar system clearly shows the cooling 
arrangement and the optics (see Fig. \ref{fig:dewarschematic}). 

\subsection{Optics}

\begin{figure}[ht]
\centering
\includegraphics[width=100mm]{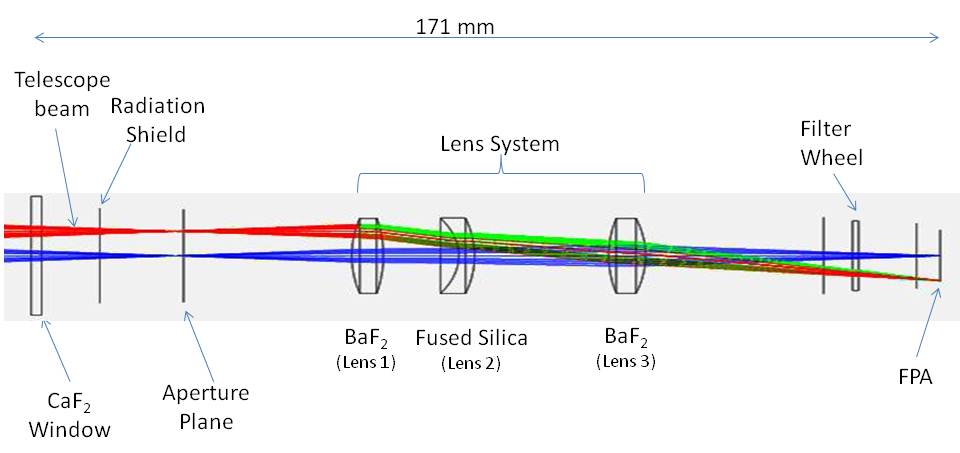}
\caption{Schematic and Zemax raytrace of TIRCAM2 optics.}
\label{fig:opticslayout}
\end{figure}

The TIRCAM2 optics consists of an arrangement of three lenses which image the
aperture plane of the photometer onto the detector plane. The design of the optics
has been optimized for the f/9 beam of the HCT while the qualifying observations were
done with the IGO telescope which has an f/10 beam. The schematic and Zemax raytrace
of the optics system are shown in Fig. \ref{fig:opticslayout}. The f/10 beam of the
IGO telescope forms an image at its focal plane which coincides with the aperture
plane of TIRCAM2. The lens system reimages the focal plane of the telescope onto the
detector plane after passing through the filter.

The lens system consists of three lenses of NIR transmitting materials,
two of which are of $BaF_{\rm2}$ and the third is of
{\it Infrared fused silica} (as shown in Fig. \ref{fig:opticslayout}), which are
at a temperature slightly higher than the second stage of the coldhead, viz., 17 deg K.
The three lenses are assembled inside an aluminium tube which
is painted matte black from the inside to reduce stray light from being scattered onto the detector.
The lenses are separated by brass spacers with a flexible spacer between lens 2 and 3 to allow
for relative movement of the lenses during cooling due to the different thermal coefficients
of expansion of the lenses and the spacers. The spacers and aluminium tube are machined at
room temperature with dimensions which are calculated to compensate for the contraction
at a temperature of 17 deg K. The gap between the lens tube and the filter wheel is covered with
aluminium tape to prevent stray light from entering the lens tube. The other end of the tube
is connected to the radiation shield by a fibreglass ring to block stray light from entering the
lens tube and to thermally isolate the lens tube which is at $\sim$17 deg K, from the radiation
shield which is at $\sim$100 deg K.

\begin{figure}[ht]
\centering
\includegraphics [width=60mm] {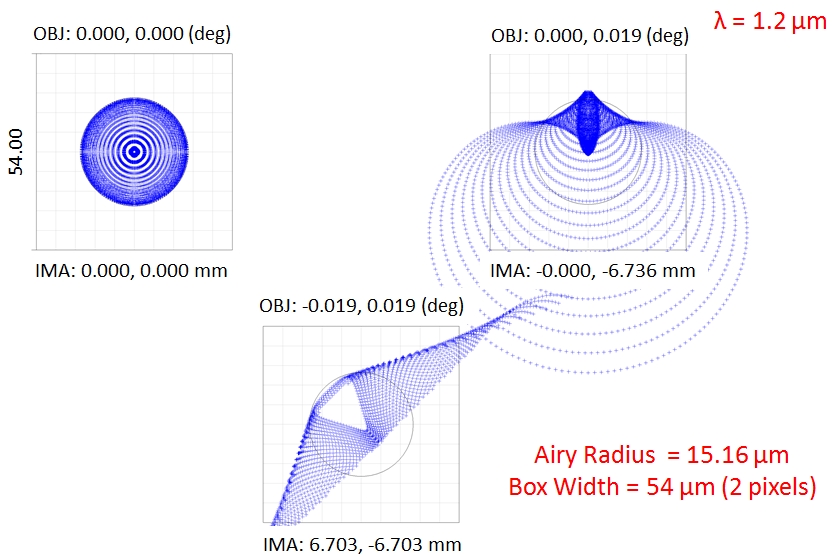}
\quad
\includegraphics [width=60mm] {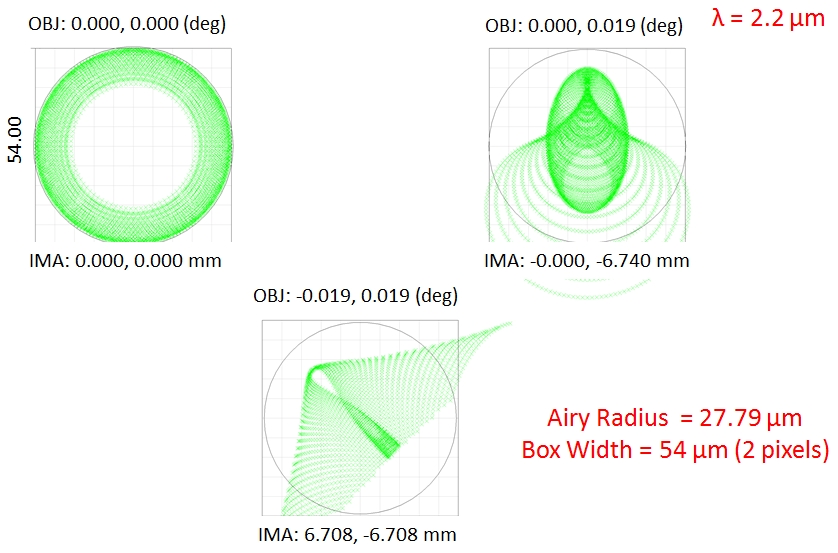}

\includegraphics [width=60mm] {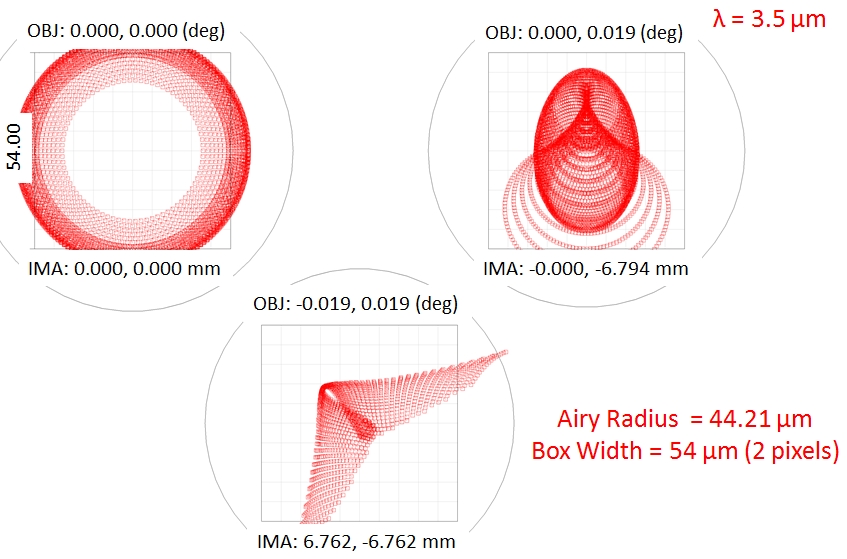}
\caption{Spot diagrams at $\lambda$ = 1.2, 2.2 and 3.5 $\mu$m ({\it clockwise from top left}).}
\label{fig:spotdiagrams}
\end{figure}

The Zemax spot diagrams are shown in Fig. \ref{fig:spotdiagrams}
at wavelengths
of 1.2, 2.2 and 3.5 $\mu$m. The three figures in each diagram
({\it clockwise from top left}) are the spot diagrams
at the centre of the detector, at a field angle of 1.14 arcmin and at
1.61 arcmin from the centre of the detector.
The pixel scale is 0.27 arcsec per pixel
and the circular unvignetted field of view (FOV) of the system is 1.14 arcmin.
The square box in the spot diagram represents 2 x 2 pixels square (1 pixel = 27 $\mu$m)
and the circle represents the Airy disk at the different wavelengths.
This shows that the image quality is close to the diffraction limit for a major part of the array.

\begin{figure}[ht]
\centering
\includegraphics [width=60mm] {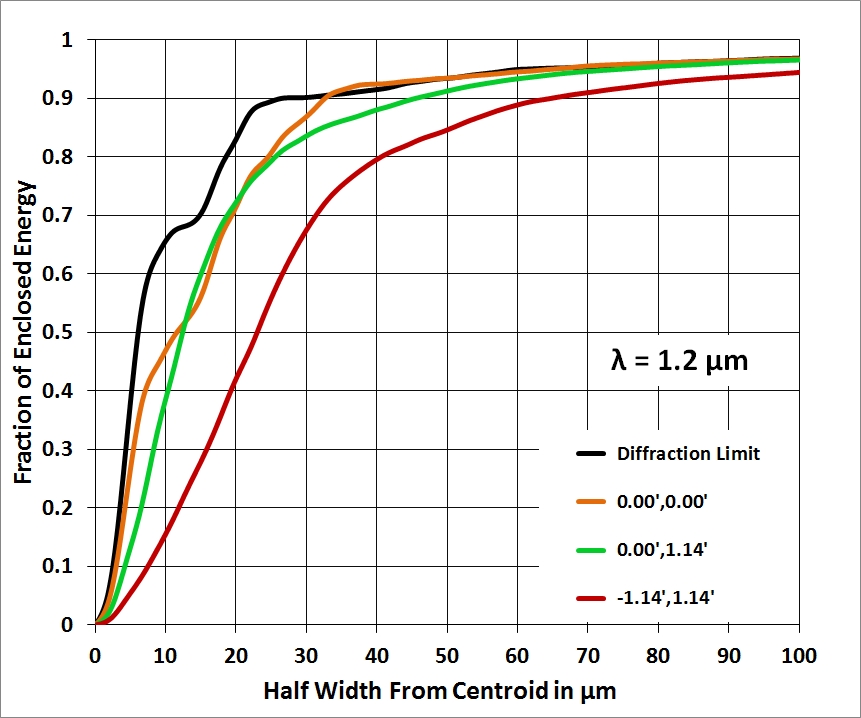}
\qquad
\includegraphics [width=60mm] {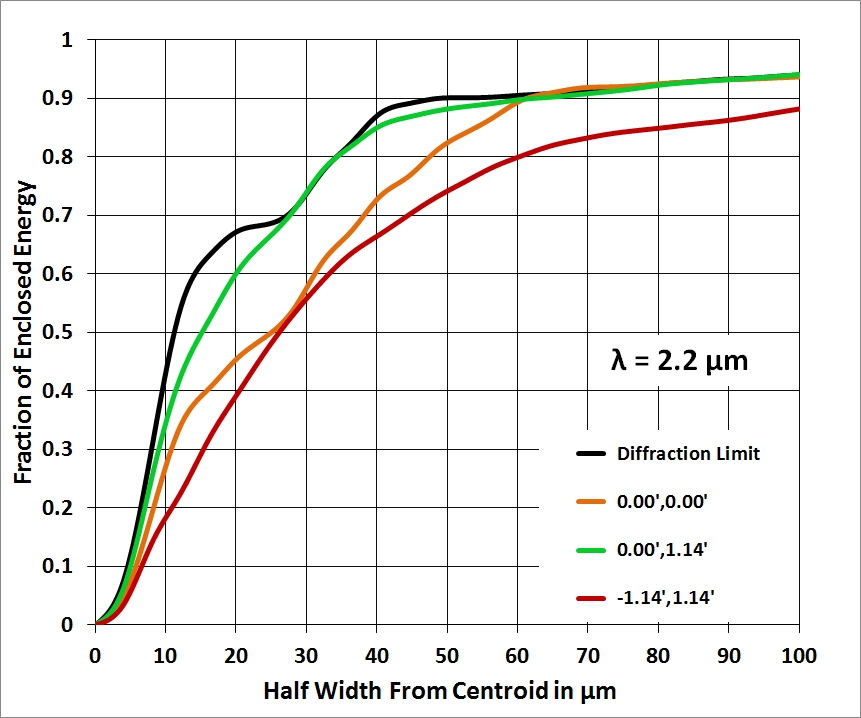}

\includegraphics [width=60mm] {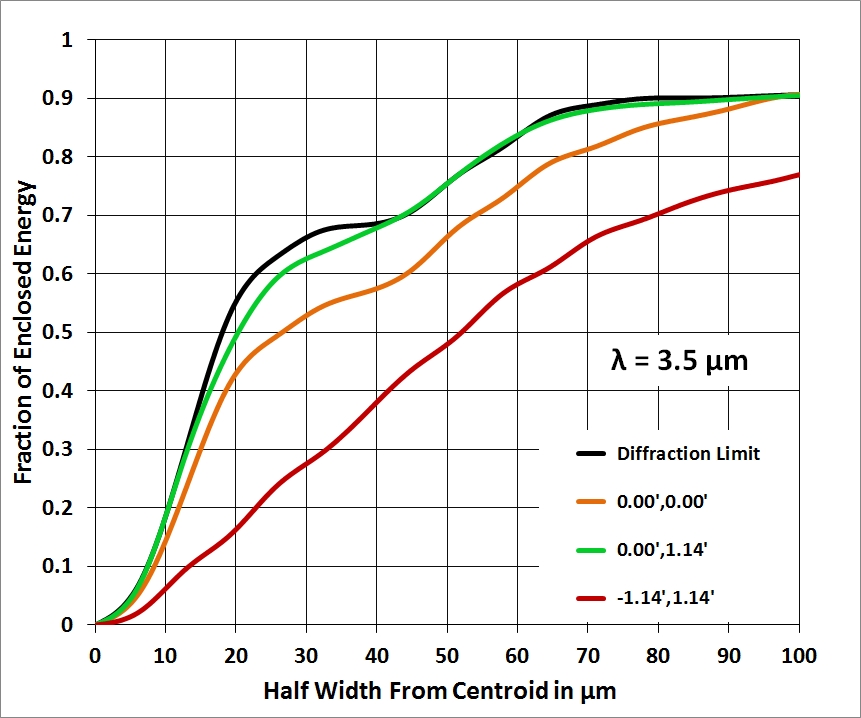}
\caption{Ensquared energy at $\lambda$ = 1.2, 2.2 and 3.5 $\mu$m ({\it clockwise from top left}).}
\label{fig:ensquareplots}
\end{figure}

Fig. \ref{fig:ensquareplots} shows the plots of the ensquared energy for different
locations on the detector at $\lambda$ = 1.2 $\mu$m,
$\lambda$ = 2.2 $\mu$m and $\lambda$ = 3.5 $\mu$m ({\it clockwise from top left}).
The black line plot represents the ensquared energy at
the diffraction limit for that particular wavelength.
The orange line plot represents the ensquared energy at the centre (0', 0'),
the green line plot represents the ensquared energy at one side (0', 1.14')
and the red line plot represents the ensquared energy at
one corner (-1.14', 1.14') of the FPA. The plots show that the image quality
is close to the diffraction limit for a major portion of the detector surface. 

\subsection{Filters}

In front of the FPA is a 8 position filter wheel with $J$, $K$, $K_{\rm cont}$, 
$H_{\rm 2}$ (1-0 S(1)), $Br\gamma$, $PAH$ (3.3 $\mu$m) and narrow L-band ($nbL$)
filters with one position blocked to enable dark frames to be captured.
The filter wheel is cooled to $\sim$17 deg K and is coupled to a
stepper motor mounted outside the dewar through a gear arrangement and a
ferrofluidic feedthrough shaft which allows the filter wheel to rotate
without degrading the vacuum of the dewar.
The position of the filter wheel is obtained from a 10-turn
potentiometer mounted on the stepper motor shaft and is used only as an
indicator. A home switch inside the dewar is used to align the filter-motor
assembly. The stepper motor can rotate a total of 8 rotations, in either
clockwise or anti-clockwise direction, and one
full rotation of the shaft is required to change to the next filter.
The filter position can be controlled to an accuracy of 0.225 degrees due to
the step size of the stepper motor and the gear ratio.

\begin{figure}[ht]
\centering
\includegraphics [scale=0.3] {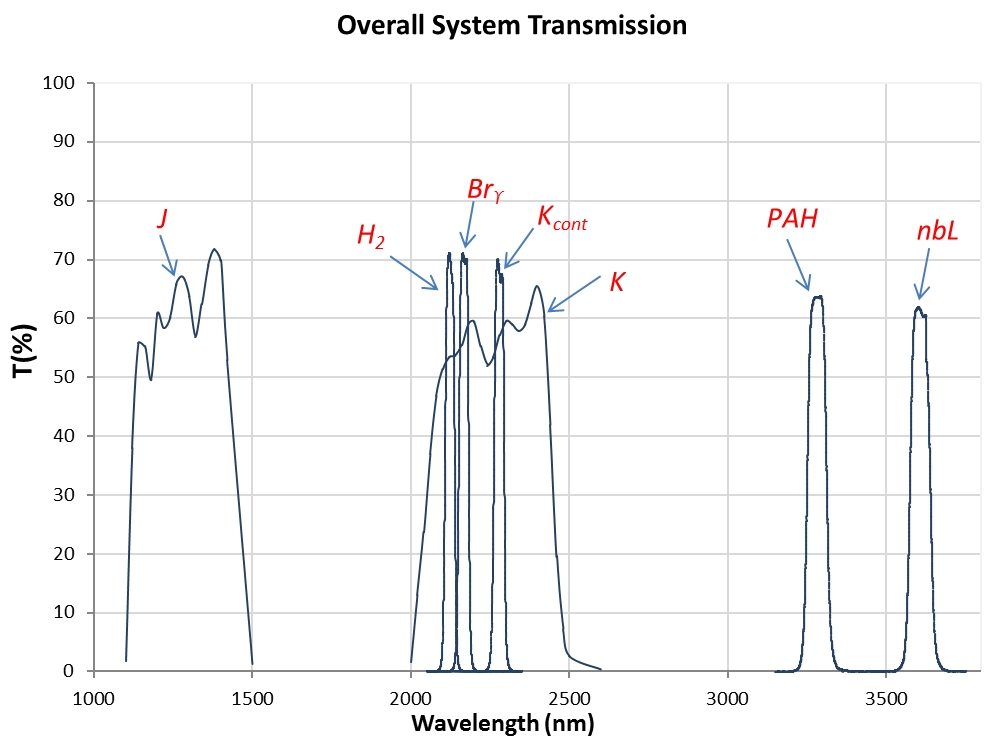}
\caption{Overall system transmission}
\label{fig:filtertransfull}
\end{figure}

Table \ref{tab:filterchar} lists the centre frequencies and the bandwidths
(FWHM) of the 7 broad- and narrow-band filters.
Fig. \ref{fig:filtertransfull} shows the overall system
transmission characteristics at 77 deg K. The shift in central wavelength due
to the operating temperature of $\sim$17 deg K and due to the change in the
angle of incidence is expected to be less than 0.03 $\mu$m in the longer
wavelength filters and is even less in the shorter wavelength filters based
on the data provided by the filter manufacturers and on the tests done on
similar filters by Stewart \& Quijada (2000).

\begin{table}[ht]
\centering
\caption{TIRCAM2 filter characteristics}
\label{tab:filterchar}
\begin{tabular}{lll}\\
\hline
Filter & $\lambda_{\rm cen}$ ($\mu$m) & $\Delta\lambda$ ($\mu$m)\\
\hline
$J$ & 1.20 & 0.36\\
$H_{\rm 2}$ & 2.12 & 0.03\\
$Br\gamma$ & 2.16 & 0.03\\
$K$ & 2.19 & 0.40\\
$K_{\rm cont}$ & 2.17 & 0.03\\
$PAH$ & 3.27 & 0.06\\
$nbL$ & 3.59 & 0.07\\
\hline
\end{tabular}
\end{table}

\subsection{Detector, Electronics and Software}

The FPA used in TIRCAM2 is an InSb, 512 rows by 512 columns,
Aladdin III Quadrant from M/s. Raytheon, USA.
This FPA has two layers: a detector layer of InSb and a
SB-206 Silicon Cryo-CMOS Read-Out Integrated Circuit (ROIC)
connected by indium bumps.
The FPA has 512 x 512 active pixels forming photodiodes, each of 
size 27 $\mu$m by 27 $\mu$m with a Source Follower per Detector
output circuit.
The total conversion gain is about 2.1 $\mu$V/electron.
The FPA requires 8 clocks and 14 dc bias voltages for a frame readout
and has 8 outputs with 8 consecutive pixels (e.g. 1-8, 9-16 ...)
being readout on any one output clock pulse. 
The pixels are read out serially to the 8 output lines by an
x-y addressing circuit that consists of two shift registers. 
The FPA pixels can be reset globally where all the pixels are reset,
or row wise where a pair of rows is reset.
The FPA requires to be cooled at a rate less than $\sim$1 deg K per minute
to allow the FPA to match volumetric changes at the detector layer
and the ROIC layer. 

The electronics of the TIRCAM2 setup comprises of the Fanout Board,
FPA controller, Filter wheel controller, 
Temperature indicator/controller and KVM (Keyboard, Video, Mouse) extender.

\begin{figure}[ht]
\centering
\includegraphics [width=100mm] {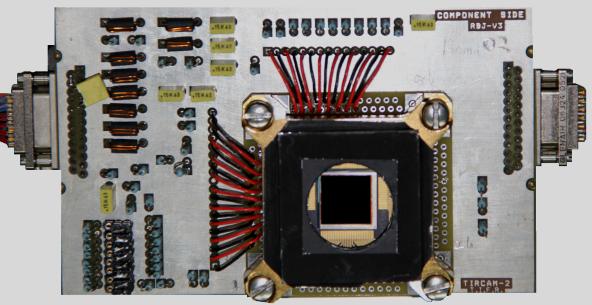}
\caption{The Fanout Board.}
\label{fig:fanoutboard}
\end{figure}

The FPA is mounted on a Fanout Board (FOB) (Fig. \ref{fig:fanoutboard})
which is designed to operate at
the low temperatures inside the dewar i.e. $\sim$35 deg K.
The FOB is a six layer PCB which is populated with LC filters for the
various biases required by the FPA and also contains the static electricity
protection circuits for the FPA inputs. The FPA is mounted on a special
socket manufactured by M/s. J. K. Henriksen \& Associates, USA.
Micro-D connectors are used to connect the biases, clocks and the
FPA outputs to the hermetic connectors mounted on the dewar body.
Fig. \ref{fig:fanoutboard} shows the populated FOB with the FPA mounted.
The biases and FPA outputs are connected to the connector on the left
whereas the clocks are connected to the connector on the right to provide
sufficient isolation between the biases and the clocks. The FOB is mounted
on a stage inside the dewar which is maintained at 35 deg K by a
resistive heater embedded in a copper block.

The FPA controller is a commercial controller from 
``Astronomical Research Camera (ARC) Inc.'' (www.astro-cam.com).
These controllers are also known as ``ARC controllers'', ``SDSU Controllers''
or ``Leach Controllers''. The controller includes  
the power supply, the controller card rack, the PCI card
for communication between the PC and the controller card rack and
the optical fibres for linking the PCI card to the controller cards.
The video board (ARC 46) is a 8 channel differential input,
16 bit digitizer with software controlled offset. It can also provide
7 bias voltages. The clock board (ARC 32) provides clocks and biases
to the FPA with  a Zener diode overvoltage protection circuit. 
The timing board (ARC 22) is a Motorola DSP 56303 based board which communicates
with the PC over the optic fibre links.  
The user interface can control parameters for the exposure time,
number of frames to capture, data filename (FITS), file storage path
and FITS header. 
The exposures can be taken either in Global (frame) reset mode or
Row pair reset mode. In Global reset mode, the full FPA is reset
(i.e. all collected charges are flushed out) followed by 
the set exposure time and the frame is then captured and digitized.
Global reset is chosen for a low background flux level.
In the row pair reset mode, two successive rows are reset followed by
the set exposure time and the row pair is then captured. In this mode
it takes 256 successive exposures to capture a single frame. The row pair
reset mode is used for high background flux levels. It takes about
256 ms to readout a frame in the Global reset mode whereas a single row pair 
is readout in 1 ms in the row pair reset mode. 

The filter wheel controller uses a microcontroller circuit
to drive a stepper motor which rotates the filter wheel and selects
the desired filter. Position feedback is obtained from a 10-turn potentiometer
mounted on the motor shaft. It takes one full rotation of the motor shaft
to move the filter wheel through one filter. With the present arrangement,
the filter wheel can be moved with an accuracy of 0.225 degrees.
The motor-filter wheel alignment is achieved by a home switch located
inside the dewar.

A commercial Lakeshore temperature controller (model 321) is used to 
control the cooling of the FPA and to maintain its temperature at 35 deg K.
The temperature sensor is a Lakeshore DT-470 diode.
A resistive heater is embedded in a copper cold finger in contact with
the FPA to provide a control over the cooling rate.
Warming up of the FPA does not require any control as the natural warm up
rate is slower than 1 deg K/minute.

A Keyboard-Video-Mouse (KVM) extender unit (Adderlink) allows the user
interface of the controlling PC unit and the controlled units to be
located several metres from each other, which is necessary in a
ground based telescope environment where the detector dewar,
control electronics and control \& storage PC are mounted on
the focal plane of the telescope, whereas the controller/operator is
situated in the telescope control room. 

\section{Cryogenics}

The detector (FPA) needs to be maintained at 35 deg K to
optimize the QE (which drops at lower temperatures), and the dark current
(which increases with higher temperature). The remaining optical components
are maintained at the coldest temperature attained by the cryocooler
system which is $\sim$17 deg K for the present configuration of the dewar
at an ambient room temperature of 22 deg C.
The TIRCAM2 dewar is evacuated to about 2 x 10$^{-6}$ mbar prior to cooling.
The pumping system is a dry turbomolecular pump of Pfeiffer Vacuum make with a capacity of
60 litres/sec.
An electrical vacuum isolation valve is always attached next to the dewar vacuum valve,
which gives safety against power failures while the dewar is cooled.
The cooling is done using a CTI make closed cycle cryocooler (Helium gas 99.999\% pure at 250 psi) 
which is a two part system with the cold head part mounted on the TIRCAM2 dewar and the compressor
part kept on the telescope floor. The two parts are connected with two flexible gas pipes 
and an electrical cable (of about 60 feet in length). The cryocooler needs to run 
continuously till the observations are completed and it requires about 2.1 KW of
continuous electrical power.
It takes about 20 hours to cool to 70 deg K from room temperature and about 1 hour
to ramp the temperature from 70 deg K to 35 deg K at the FPA.

\section{TIRCAM2 Observational Setup}

\begin{figure}[ht]
\centering
\includegraphics [width=120mm] {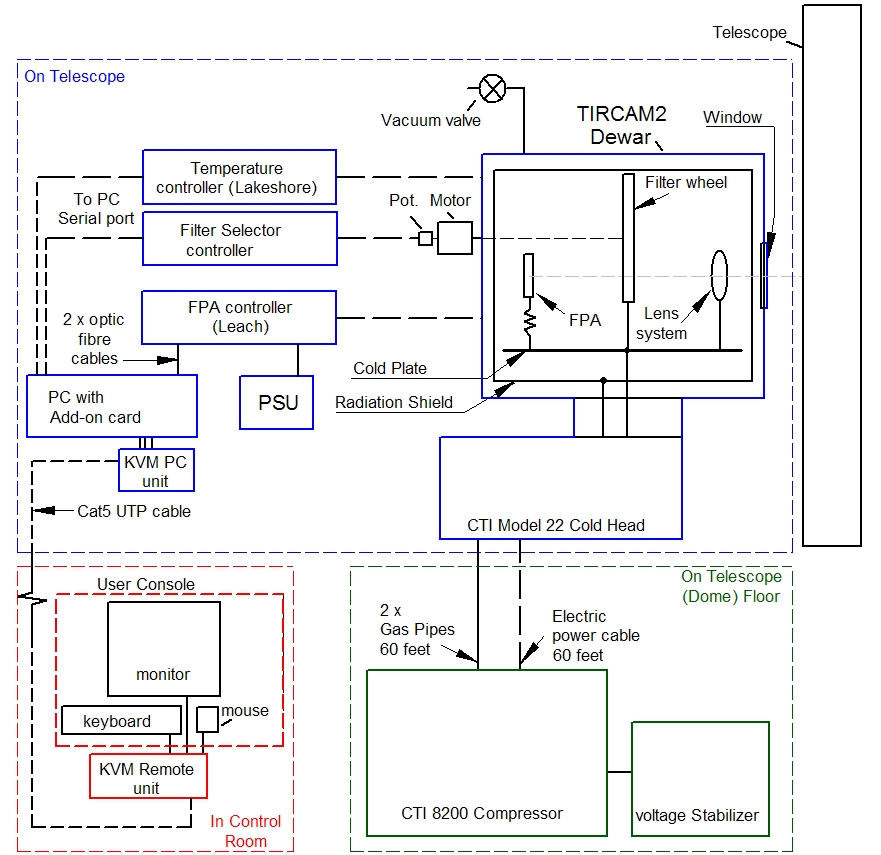}
\caption{Layout of the TIRCAM2 observational setup.}
\label{fig:Obssetup}
\end{figure}

Fig. \ref{fig:Obssetup} shows the observational setup of TIRCAM2, which shows the TIRCAM2 dewar
along with the FPA controller, temperature controller, filter wheel controller, crycooler
compressor and data acquisition computer. The user console of the data acquisition computer
is placed in the control room. The envelope of TIRCAM2 is about 100 cm
x 73 cm x 65 cm (Height x  Width x Breadth) and the mass is about 65 Kg.
During mounting of the TIRCAM2 on the telescope focal plane, a hydraulic lift table is used
to simplify the alignment of the TIRCAM2 mounting flange with the telescope port flange.

\section{Observations, Data Reduction and Performance}

\begin{figure}[ht]
\centering
\includegraphics [width=120mm] {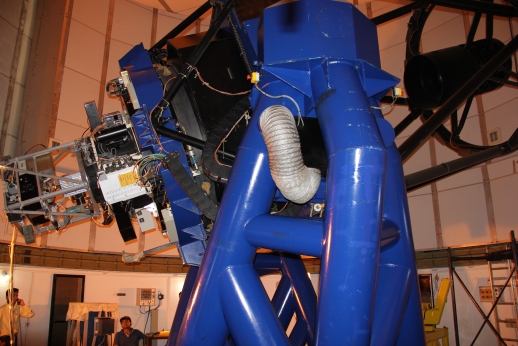}
\caption{Observational setup of TIRCAM2 at the 2-m IGO telescope.}
\label{fig:IGO}
\end{figure}

TIRCAM2 had its engineering and science runs with the 2-m IGO telescope which is located
at Girawali, about 80 kms from Pune (India), on a hill 1000m above mean sea level,
during February, March and December 2011 (Ojha et al. 2012).  
The observations were 
mainly performed during the bright sky (near full moon period) in 
February - December 2011, using the TIRCAM2 system at the f/10 Cassegrain focus of the
2-m IGO telescope. Fig. \ref{fig:IGO} shows the
TIRCAM2 system mounted at the direct port of the IGO telescope. The dark current measured
was 12 electrons/sec and the readout noise was $\sim$30 electrons as
compared to the datasheet value of 10 to 50 electron noise for the FPA.
The observations were also performed during the bright sky (near full moon period) in
May 2012, using TIRCAM2 system at f/13 Cassegrain focus of the 1.2-m telescope
of Gurusikhar Observatory at Mount Abu (altitude $\sim$ 1722m). 

We have observed several bright infrared standard sources, the Trapezium cluster in the Orion
region, McNeil's nebula (Ninan et al. 2012) and a 
few galaxies in the $J$, $K$ and $nbL$ bands during
the observational runs. The typical seeing
was 1 - 2 arcsec during the observations. We obtained several dithered exposures of
the targets in each of the NIR bands. Typical integration times per frame were
10, 0.2 and 0.02 s in the $J$, $K$ and $nbL$ bands, respectively. The images were co-added
to obtain the final image in each band. We also obtained several dithered sky 
frames close to the target position in each NIR band. Photometric calibration 
was done from the observations made on the same nights, on the United 
Kingdom Infrared Telescope (UKIRT) standard stars, 
at airmasses close to that of the target observations. The images obtained by TIRCAM2 
were written in FITS format and were processed with IDL and IRAF scripts. 
All the NIR images went through the standard pipeline 
reduction procedures like electronic gain correction, bad pixel masking \& correction,
dark/sky-subtraction, flat-fielding, co-adding images, and magnitude calculation. 
Photometric magnitudes 
were extracted using the iraf daophot/phot and apphot/phot tasks.

\begin{figure}[ht]
\centering
\includegraphics [width=33mm] {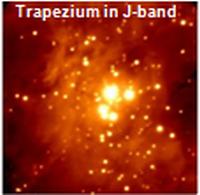}
\qquad
\includegraphics [width=33mm] {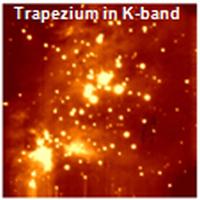}
\qquad
\includegraphics [width=33mm] {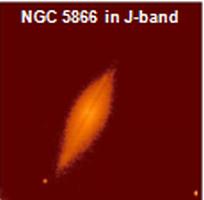}

\includegraphics [width=33mm] {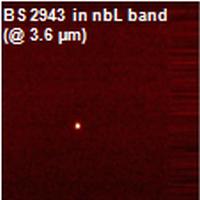}
\qquad
\includegraphics [width=33mm] {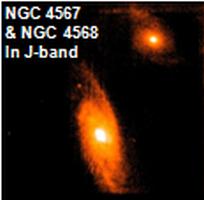}
\qquad
\includegraphics [width=33mm] {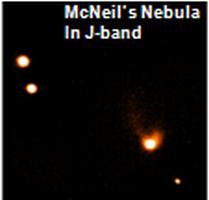}
\caption{The sample images ($\sim$2.3 arcmin x 2.3 arcmin) taken using TIRCAM2 mounted on the 2-m IGO
telescope.}
\label{fig:images}
\end{figure}

Fig. \ref{fig:images} shows the sample images ({\it clockwise from top left}: Trapezium cluster
in the $J$-band and in the $K$-band, NGC 5866 lenticular galaxy, McNeil's nebula,
NGC 4567 \& NGC 4568 twin galaxies and BS 2943 star)
taken with TIRCAM2 using the 2-m IGO telescope. 

\begin{figure}[ht]
\centering
\includegraphics [width=40mm] {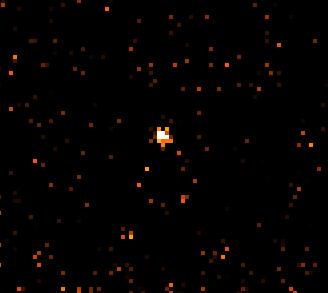}
\caption{The faintest star (BS 2721) observed in the $nbL$-band with a total integration 
time of 0.5s. Its $L$-band magnitude is 4.08.}
\label{fig:BS2721}
\end{figure}

The limiting magnitude obtained from
the analysis of the Trapezium cluster field is 16.3 (T$_{\rm int}$ $\sim$ 1050s) 
and 14.5 mag (T$_{\rm int}$ $\sim$ 164s) in the $J$ and $K$ bands, respectively. 
In the $nbL$-band, the faintest object we observed is BS 2721 from the IGO 
on 2011 February 17, having an $L$-band 
magnitude of 4.08 (T$_{\rm int}$ $\sim$ 0.5s). 
Fig. \ref{fig:BS2721} shows the image obtained by combining 25 dithered frames.
Typical integration time per frame was 0.02s. 

\begin{figure}[ht]
\centering
\includegraphics [width=120mm] {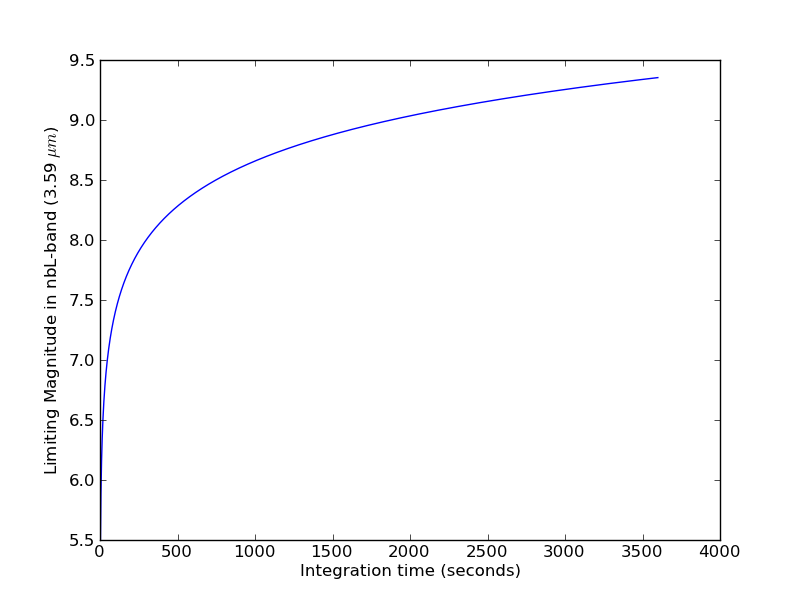}
\caption{The limiting magnitude with $5 \sigma$ detection criteria in the 
$nbL$-band (3.59 $\mu$m) as a function of exposure time in seconds. The atmospheric conditions 
are assumed to be similar to 2011 February 17 at IGO.}
\label{fig:limitMag_exp}
\end{figure}

Assuming a similar sky condition, we estimated the limiting magnitude in the 
$nbL$-band for a longer exposure time. A limiting magnitude was taken to be 
the magnitude of the star which has its peak of flux profile at 5 $\sigma$ 
level above background, where $\sigma$ is the background noise. From our BS 2721 
image we obtained the sky $\sigma$ $\sim$ 0.40 ADUs for a 0.1s dithered frame. Combining 
5 such frames reduced the $\sigma$ to $\approx$ 0.16 ADUs. As expected, this is 
indeed reducing as $\sqrt{N}$. Assuming the same sky condition, we estimated 
the flux ratio of a limiting magnitude star with respect to BS 2721. Combining 
this with the $\frac{1}{\sqrt{t}}$ fall of the background standard deviation 
$\sigma$, we get the limiting magnitude $M_{limit} = 4.08 + 0.828 + 1.25\log{t}$ .
Note that, this is valid for the same atmospheric condition as of 2011 February 17 
at IGO (FWHM $\sim$ 1 arcsec). Using the above formula, if we observe for 1 hour then the expected 
limiting magnitude comes out to be 9.3 in the $nbL$-band.
Fig. \ref{fig:limitMag_exp} shows the plot of limiting magnitude in the $nbL$-band 
{\it versus} total exposure time.

Our limiting magnitude is determined by the background $\sigma$. If
the background flux reduces by $X$ magnitude then the limiting magnitude will 
increase by $\frac{X}{2}$ magnitude. 

\begin{figure}[ht]
\centering
\includegraphics [width=100mm] {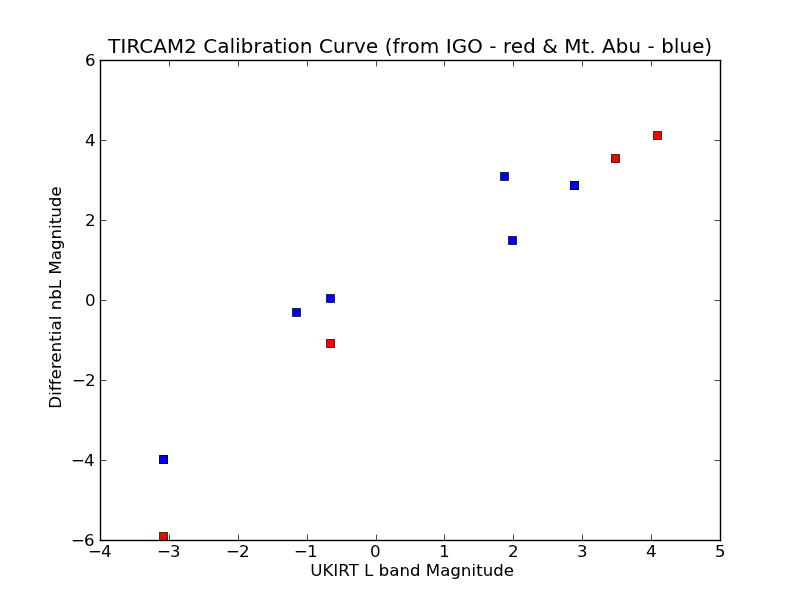}
\caption{Instrumental magnitudes from log of ADUs/sec {\it versus} 
actual $L$-band magnitudes from UKIRT, showing linearity of TIRCAM2 in the $nbL$-band 
(red squares -- IGO data, blue squares -- Mount Abu data).}
\label{fig:LinearityTIRCAM2}
\end{figure}

To check the linearity of counts in the $nbL$-band, we took the data of those 
nights where 3 or more standard sources were observed. The log of observations
is shown in Table 2. 
Fig. \ref{fig:LinearityTIRCAM2} shows the plot of instrumental magnitudes 
calculated using log of ADUs/sec {\it versus} standard $L$-band magnitudes from
the UKIRT. The plot shows that our array is linear in this regime. 

It is interesting to compare the {\it Spitzer}-IRAC values of saturation limit
in the 3.6 $\mu$m band. For a frame time of 2s, the point source saturation limit 
in the IRAC 3.6 $\mu$m band is $< 7.92$ mag. TIRCAM2 can therefore be 
used to observe the range of magnitudes brighter than the saturation limit of 
{\it Spitzer}-IRAC.

\begin{table}
\caption[]{Log of standard star observations in the $nbL$-band. Observations were 
carried out in 2011 and 2012 from IGO and Mount Abu, respectively.}
\vskip 0.1cm
\centering
\begin{tabular}{|c|c|c|c|}
\hline
Date & Object & $L$-band mag from UKIRT & Total integration time  \\
\hline
17/Feb/2011 & BS2943 & -0.66 & 0.5s \\
 & BS2721 &  4.08  & 0.5s \\
 & BS4983 &  2.88  & 0.5s \\

18/Feb/2011 & BS4983 &  2.88  & 1s \\
 & BS5340 & -3.09  & 0.2s \\

20/Feb/2011 & BS5340 & -3.09  & 0.56s \\
 & BS4983 &  2.88  & 0.5s \\
 & BS5447 &  3.48  & 0.5s \\
\hline
04/May/2012 & BS3903  & 1.98  & 0.14s \\
 & BS4983 &  2.88 & 0.7s \\
 & BS2990 & -1.15  &  0.2s \\

06/May/2012 & BS2943 & -0.66 & 0.08s \\
 & BS5340 & -3.09 & 0.12s \\
 & BS6136 &  1.86 & 0.7s \\

\hline
\end{tabular}
\label{table:log_OBS}
\end{table} 

\section{Conclusions}

The upgradation of the near infrared camera TIRCAM2 to utilize a 512 x 512 InSb array
and additional narrow-band filters for use at the focal plane of the Indian 2-m class telescopes
was successfully carried out. The systems were tested and the observing runs of TIRCAM2
at the IUCAA 2-m telescope at Girawali and PRL 1.2-m telescope at Gurusikhar were quite successful
even though the TIRCAM2 observations were made during the near full moon period.
We could also observe sources in the $nbL$-band ($\sim$ 3.6 $\mu$m) from the Girawali 
(altitude $\sim$ 1000m) and Gurusikhar (altitude $\sim$ 1722m) sites. 
Our limiting magnitude estimate shows that TIRCAM2 can be used to observe the 
bright magnitude range at 3.6 $\mu$m below {\it Spitzer-IRAC} saturation limit.
In the near future, with a few modifications in the optics, we plan to explore 
TIRCAM2's performance in the $PAH$ (3.3 $\mu$m) and
$M$ (4.5 $\mu$m) bands from the Hanle (altitude $\sim$ 4500m) site.

\bigskip
\bigskip
\noindent
{\bf Acknowledgments}

\vskip .2cm
We thank Prof. A.N. Ramaprakash and Dr. V. Mohan of Inter-University Centre for Astronomy \&
Astrophysics (IUCAA, Pune) for their valuable assistance in interfacing
the TIRCAM2 to the IGO telescope.
We also thank the staff of IGO, operated by IUCAA, for their assistance and support 
during the observations.
We thank Prof N.M. Ashok and the staff of Gurushikar Observatory, operated by PRL,
for their assistance and support during the observations.
We are grateful to M/s. Infrared Laboratories, M/s. Raytheon and Dr. Bob Leach
(M/s. Astronomical Research Camera Inc.)
for their help in realising this project.
We also thank Prof. S.N. Tandon for his valuable comments and suggestions
which have helped us to improve the contents of this paper.

\label{lastpage}

\end{document}